\documentclass[aps,prc,twocolumn,superscriptaddress,showpacs,floatfix,nofootinbib]{revtex4-1}
\usepackage{amsmath,graphicx}

\begin{document}

\title{Transverse momentum structure of pair correlations as a signature of collective behavior in small collision systems}

\author{Igor Kozlov}
\affiliation{
McGill University,  3600 University Street, Montreal QC H3A 2TS, Canada}
\author{Matthew Luzum}
\affiliation{
McGill University,  3600 University Street, Montreal QC H3A 2TS, Canada}
\affiliation{
Lawrence Berkeley National Laboratory, Berkeley, CA 94720, USA}
\author{Gabriel Denicol}
\author{Sangyong Jeon}
\affiliation{
McGill University,  3600 University Street, Montreal QC H3A 2TS, Canada}
\author{Charles Gale}
\affiliation{
McGill University,  3600 University Street, Montreal QC H3A 2TS, Canada}
\affiliation{
Frankfurt Institute for Advanced Studies, Ruth-Moufang-Str. 1, D-60438 Frankfurt am Main, Germany}
\date{\today}

\begin{abstract}
We perform 3+1D viscous
hydrodynamic calculations of proton-lead and lead-lead collisions at top LHC energy.  
We show that existing data from high-multiplicity p-Pb events can be well described in
hydrodynamics, suggesting that collective flow is plausible as a correct description of these collisions.  
However, a more stringent test of the presence of
hydrodynamic behavior can be made by studying the detailed momentum
dependence of two-particle correlations.  We define a relevant observable, $r_n$, and make
predictions for its value and centrality dependence if hydrodynamics
is a valid description.  This will provide a non-trivial confirmation
of the nature of the correlations seen in small collision systems, and
potentially to determine where the hydrodynamic description, if valid anywhere, stops
being valid.  Lastly, we probe what can be learned from this observable, finding 
that it is insensitive to viscosity, but sensitive to aspects of the initial state of the system
that other observables are insensitive to, such as the transverse length scale of the
fluctuations in the initial stages of the collision.
 \end{abstract}

\maketitle

\section{Introduction}
It is believed that ultrarelativistic collisions between heavy nuclei
produce an almost-perfect fluid Quark-Gluon Plasma.  Some of the
strongest evidence for this conclusion comes from correlations between
hadrons that are emitted from the collision region, and in particular
the strong azimuthal dependence.
Recent measurements from collisions of protons and deuterons 
with heavy nuclei reveal that high multiplicity events display features 
that are strikingly similar to those previously seen in heavy-ion collisions 
\cite{CMS:2012qk, Abelev:2012ola, Aad:2012gla, Adare:2013piz, Wang:2014qiw}.
This raises the question of whether these small
systems also behave hydrodynamically,  whether the physics governing the 
two systems is different and the similarity in
observed correlations is merely a coincidence, or whether the data from
heavy-ion collisions have been misinterpreted.  Various explanations
for the observed correlations in these small systems have been proposed
\cite{Dusling:2013oia}, 
but a definitive answer is not yet available.

In this paper we perform 3+1D event-by-event viscous hydrodynamic calculations 
of p-Pb collisions at 5.02 TeV,
in order to determine whether existing data is consistent with a fluid medium, 
and whether new measurements can be devised in order to provide a definitive
resolution to the question of the correct description of the system.
\section{Hydrodynamic calculation}
We model the evolution of the collision system with the 3+1D relativistic viscous hydrodynamics solver MUSIC version 2.0.  Details can be found in Ref.~\cite{Schenke:2010nt}. 
In all calculations, we use the following parameters, which have given reasonable fits to heavy-ion data in the past:  thermalization time $\tau_0$ = 0.6 fm/$c$, freeze out temperature $T_{\rm freeze}$ = 150 MeV, and equation of state s95p-v1 \cite{Huovinen:2009yb}.

For initial conditions, we use a modified Monte Carlo Glauber model, where a contribution of entropy density is associated with each participating nucleon.  The simplest prescription is to distribute entropy in the transverse plane according to a 2D Gaussian centered at the location of each participant:  
\begin{equation}
\label{eq:trans}
\rho_{\perp} (\vec{x}_\perp) \equiv \frac{1}{2\pi\sigma^2}  \exp\left(-\frac{|\vec{x}_\perp|^2}{2\sigma^2}\right) .
\end{equation}
The width of the transverse Gaussian $\sigma$ is commonly chosen to be between 0.4 fm and 0.8 fm.  We will vary the value within this range in order to study the effect of the granularity of the initial state and the transverse length scale associated with density fluctuations in the transverse plane. 

\begin{figure}
 \includegraphics[width=\linewidth]{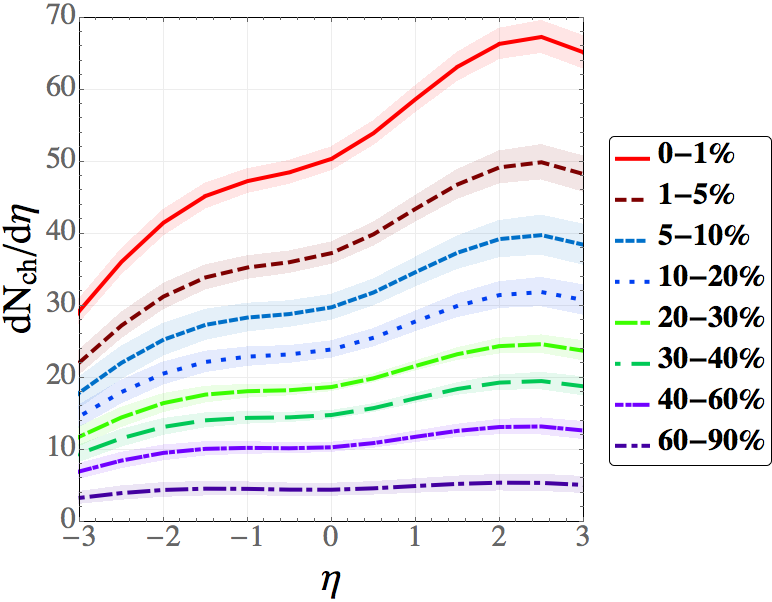}
 \caption{ (Color online) 
Charged hadron multiplicity $dN/d\eta$ versus pseudorapidity in our calculation for various centrality bins, to be compared to preliminary data from the ATLAS collaboration shown in Fig. 2 of Ref.~\cite{AlexanderMilovonbehalfoftheATLAS:2014rta}.}  
\label{fig:dNdeta}
\end{figure}

For the distribution in spatial rapidity, one should take into account the fact that a proton-nucleus collision is not symmetric.  In particular, there are more particles produced in the direction of the nucleus, and the asymmetry is greater in events with a larger multiplicity (see, e.g., Fig.~\ref{fig:dNdeta}).  This can be achieved by associating an asymmetric rapidity profile with each participant, peaked in the direction of its motion.  Following Ref.~\cite{Bozek:2010bi}, we take the profile
\begin{equation}
\rho_{L\pm} (\eta) \equiv \left( 1 \pm \frac \eta {y_{\rm beam}} \right) \exp\left[ -\frac{\left(|\eta| - \eta_0 \right)^2}{2\sigma_\eta^2} \theta\left(|\eta| - \eta_0 \right)\right] ,
\end{equation}
with parameters $\eta_0$ = 2.5, $\sigma_\eta$ = 1.4, and $y_{\rm beam}$ is the beam rapidity 8.58.
So right moving participants have a contribution proportional to $\rho_{L+}$ while left moving participants have a contribution proportional to $\rho_{L-}$.  Events with more participants in one direction will then naturally have an asymmetry.  

\begin{figure}
 \includegraphics[width=\linewidth]{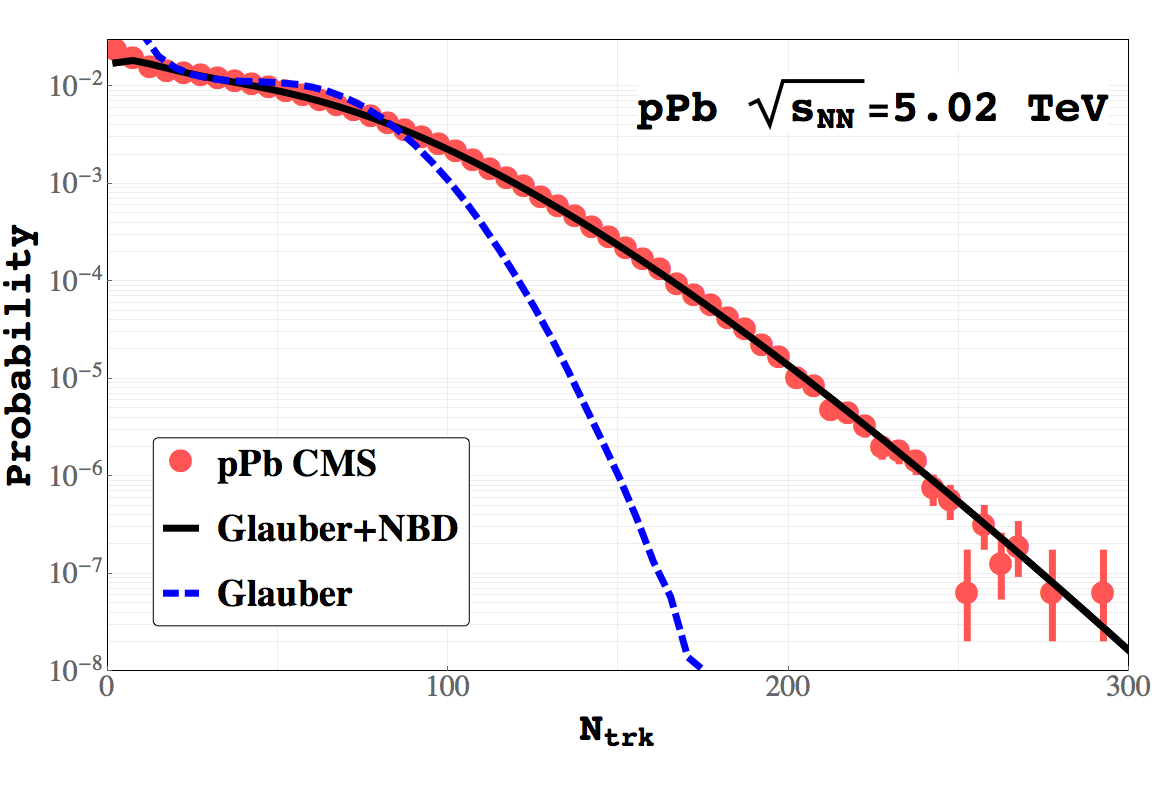}
 \caption{ (Color online) 
Distribution of (uncorrected) multiplicity $N_{\rm trk}$ at $|\eta| < 2.4$ and $p_T > 0.4$ GeV in p-Pb events from the CMS Collaboration \cite{Chatrchyan:2013nka} compared to the entropy distribution implied by a basic MC Glauber model with a fixed entropy per participant, and the model used in this work that has been supplemented with additional negative binomial fluctuations (Glauber + NBD).  }  
\label{fig:mult}
\end{figure}

\begin{figure}
 \includegraphics[width=\linewidth]{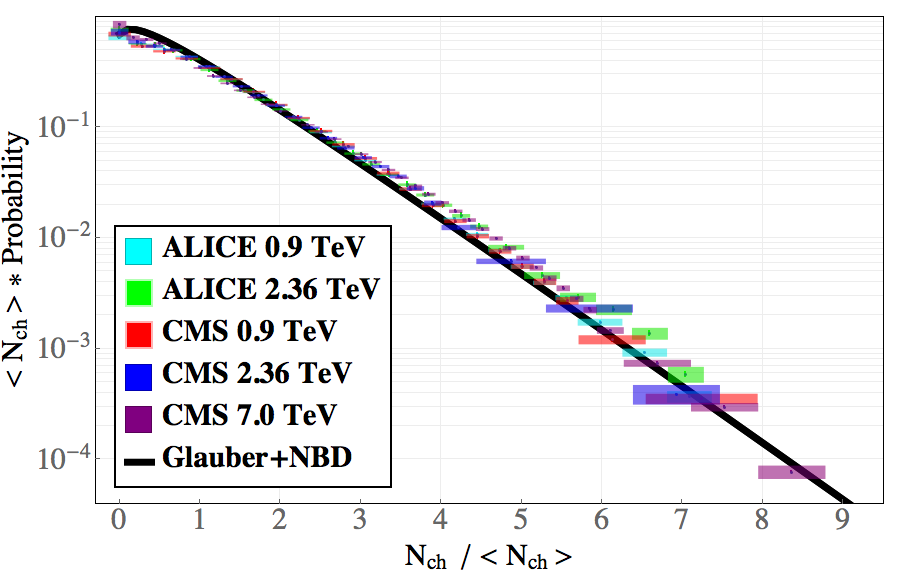}
 \caption{ (Color online) 
Scaled NSD charged hadron multiplicity distributions $\langle N\rangle P(N/\langle N\rangle)$ at $|\eta| < 0.5$ in proton-proton collisions at various collision energies from the ALICE Collaboration \cite{Aamodt:2010ft}, and the CMS Collaboration \cite{Khachatryan:2010nk}, compared to the distribution of entropy given by our model.  Data are not available for the same collision energy as the p-Pb collisions that our model is tuned to, but normalizing by $\langle N \rangle$ gives a universal curve that is reproduced for a range of collision energies (`KNO scaling').
}  
\label{fig:ppmult}
\end{figure}

The total initial entropy density distribution is then given as a sum over participants
\begin{equation}
s(\vec{x}_\perp, \eta, \tau = \tau_0) = \sum_{i=1}^{N_{\rm part}} s_i\ \rho_{\perp} (\vec{x}_\perp - \vec{x}_{i})\ \rho_{L\pm} (\eta) .
\label{NBD}
\end{equation}
Here $\vec x_i$ is the transverse position of each participant nucleon and $s_i$ is the entropy per participant (per unit rapidity at $\eta=0$).  Often this is taken to be a constant.  However, this is not realistic.  It implies, for example, that in the limit of a proton-proton system, every collision will produce the same multiplicity.  On the contrary, it is known that proton-proton collisions exhibit a wide distribution of multiplicities, with a long tail, which is well described by a negative binomial distribution.  Similarly, this basic Glauber model implies a multiplicity distribution in a p-Pb system that is much narrower than seen experimentally (see Fig.~\ref{fig:mult}).  However, instead of a constant, we can sample the factor $s_i$ for each participant according to a negative binomial distribution 
\begin{equation}
P (s_i) = \frac {\Gamma(s_i + s_0\kappa) (s_0\lambda)^{s_i} (s_0\kappa)^{s_0\kappa}} {\Gamma(s_0\kappa) s_i ! (s_0\lambda + s_0\kappa)^{s_i + s_0\kappa}}.
\end{equation}
In this case, the mean entropy per participant is $\langle s_i \rangle = s_0 \lambda$.  If  we choose parameters $\lambda=5.11$ and $\kappa=0.62$, the scaled entropy distribution (i.e., with inverse multiplicity per unit entropy $s_0\to1$) approximately fits both p-p (Fig.~\ref{fig:ppmult}) and p-Pb data (Fig.~\ref{fig:mult}).   Since the initial entropy is approximately proportional to the final multiplicity in each event, this will result in an approximately correct distribution of multiplicity, when scaled by the proper factor $s_0$.  Note that it is particularly important for this work to have a realistic description of the tail of the multiplicity distribution, since the events with the very highest multiplicity are the best candidates for a hydrodynamic description.
%

Finally, in order to facilitate comparisons to experiment, in the p-Pb case we shift from the frame representing the center of mass of an individual nucleon-nucleon collision to the experimental lab frame, which is shifted by a rapidity of 0.465 in the direction of the lead beam.

With this prescription, we can generate a very large number of initial conditions and place them into ``centrality'' bins according to their total entropy.  By calculating the multiplicity per unit entropy in a set of high-multiplicity events, we can then choose $s_0$ in Eq.~\eqref{NBD} so that the measured multiplicity in the most central bin (representing a fraction $6\times 10^{-5}$ of events) matches experiment.
(This is done separately for each set of parameters, since the amount of entropy produced in the collision can vary.)  
After this has been done, we prepare events in the same multiplicity (entropy) bins in a Pb-Pb system, using exactly the same model parameters, for comparison.  Once a sufficient set of initial conditions are prepared, we evolve them with hydrodynamics and calculate the distribution of charged hadrons in each event, parameterized as:

\begin{equation}
\label{vn}
   \frac {dN} {dp_T d\eta d\phi} \equiv \frac 1 {2\pi}   \frac {dN} {dp_T d\eta}\left[1+ \sum_{n=1}^\infty v_n \cos n(\phi - \Psi_n)\right] , 
\end{equation}
where in general $v_n$ and $\Psi_n$ depend on pseudorapidity $\eta$ and transverse momentum $p_T$.  In a purely hydrodynamic picture, this single-particle distribution contains all possible information, and observables can be calculated from appropriate event averages, as described in the following section.

%



\begin{table}
\begin{tabular}{l l l l}
Centrality & Fraction & $\langle N_{\rm ch}\rangle$ & CMS $\langle N_{\rm trk}\rangle$ \\
\hline 
0.0000031-0.0000631&0.00006&280&280\\
0.0000631-0.0005631&0.0005&230&236\\
0.0005631-0.0045631&0.004&190&195\\
0.0045631-0.02&0.0154369&150&159\\
0.02-0.05&0.03&120&132\\
0.05-0.12&0.07&95&108\\
0.12-0.24&0.12&70&84\\
0.24-0.33&0.09&55&66\\
0.33-0.43&0.1&45&54\\
0.43-0.55&0.12&35&42\\
0.55-0.69&0.14&20&30\\
0.69-1.&0.31&8&12\\
\end{tabular}
\caption{
Centrality bins used for the hydrodynamic calculations.  The p-Pb hydro events were selected according to the total initial entropy, in bins corresponding to the fraction of the cross section listed in the first column.  The results can then be compared directly to data selected according to multiplicity in bins with the same fraction of the cross section \cite{Chatrchyan:2013nka}, or rebinned for comparison to other centralities (as in Fig.~\ref{fig:dNdeta}).  The last column lists the (uncorrected) number of tracks from the respective CMS measurements in Ref.~\cite{Chatrchyan:2013nka}, to which we map our results when comparing to their data.   Calculations with the same cuts in entropy were then performed for Pb-Pb events, to be compared to experimental measurements in the same multiplicity bins.
\label{tab}
}
\end{table}

\section{Comparison to Experiment}
The first task is to determine whether a hydrodynamic calculation, with realistic 
properties, can describe measured data.  If so, it is confirmed as a plausible
explanation.   A number of hydrodynamic calculations are available 
\cite{Bozek:2011if, Bozek:2012gr, Bozek:2013uha, Bzdak:2013zma, Qin:2013bha, Schenke:2014zha}. 
Here, we combine all the relevant existing data and extend far into the high-multiplicity tail.
In addition, we explore parameter space that has not been studied.

The simplest observables (theoretically) are single-particle measurements.  
For example, one can measure the average number of charged hadrons 
in each bin in pseudorapidity.  Our results are shown in Fig.~\ref{fig:dNdeta},
showing that our prescription
for the longitudinal profile is reasonable.

\begin{figure}
 \includegraphics[width=\linewidth]{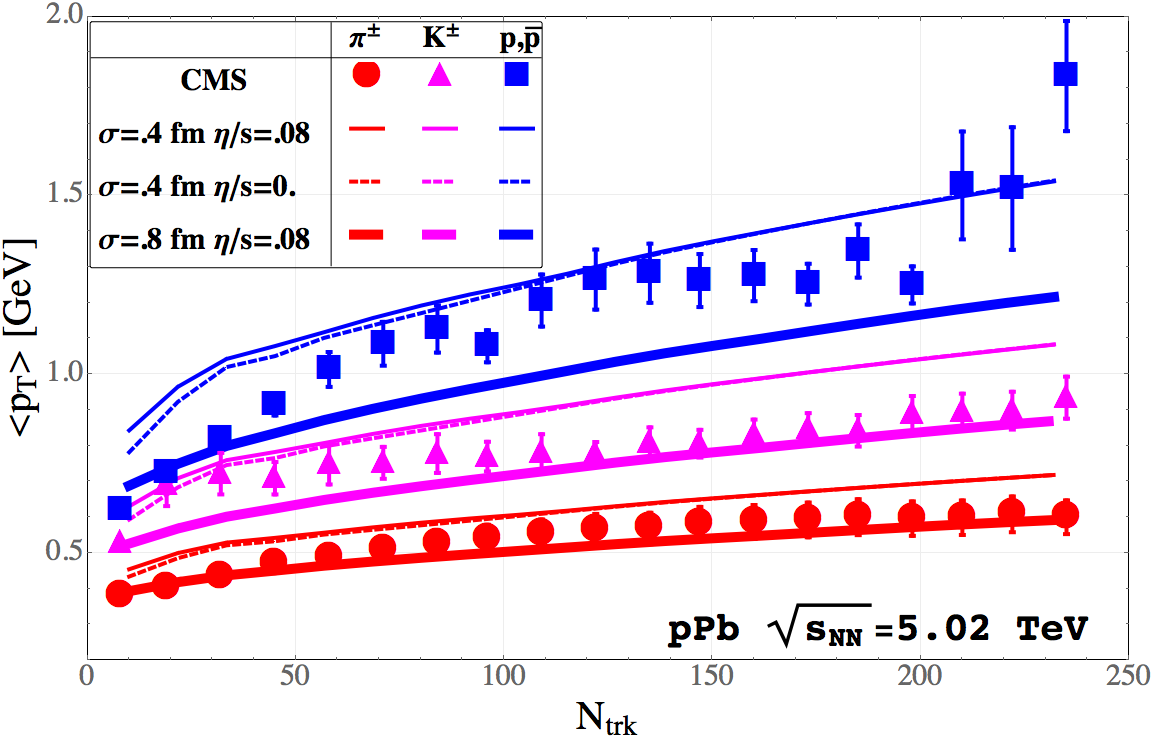}
 \caption{ (Color online) 
Average transverse momentum of identified particles in 5.02 GeV p-Pb collisions compared to CMS data \cite{Chatrchyan:2013eya}.}  
\label{fig:meanpt}
\end{figure}

Next we compare to the mean transverse momentum for identified particles, in Fig.~\ref{fig:meanpt}.  A smaller value of $\sigma$ (recall Eq.~\ref{eq:trans}) corresponds to a more granular initial condition as well as a smaller overall transverse size of the system.  The larger gradients in the initial condition then result in a larger  average transverse momentum, while shear viscosity has little affect.  From these results it is clear that these data can be well described by a hydrodynamic calculation, at least for high multiplicity events.  
(Note that most of the x-axis represents a small fraction of events.  See Table \ref{tab}.)
Note also that parameters such as the thermalization time and freeze out temperature were held fixed.  For precise numerical analysis (e.g., extracting the preferred value of $\eta/s$), they should be adjusted to fit this $p_T$ spectra for each set of parameters to be investigated, but for the purposes of this work, it is  unnecessary.

The most striking aspect of the data, however, is the strong azimuthal dependence.  To study this it is necessary to measure the correlation between two or more particles.  The simplest possibility is the distribution of pairs of particles.  The dependence on relative azimuth $\Delta\phi = \phi_1 - \phi_2$ can be captured by Fourier components.  We can define the observable:
\begin{equation}
V_{n\Delta} (p_T^a, p_T^b) \equiv \left\langle \frac 1 {N_{\rm pairs}} \sum_{\rm pairs} \cos n\Delta\phi \right\rangle .
\end{equation}
Here and in the following, angular brackets represent a simple average over events $\langle\ldots\rangle\equiv (\sum_{\rm events} \ldots)/ N_{\rm events}$. The labels $a$ and $b$ refer to the transverse momentum bin from which each particle in the pair is selected.   Note that, in principle, the particles can be selected from different bins in pseudorapidity as well as $p_T$, but here we focus on the transverse momentum dependence.  

\begin{figure}
 \includegraphics[width=\linewidth]{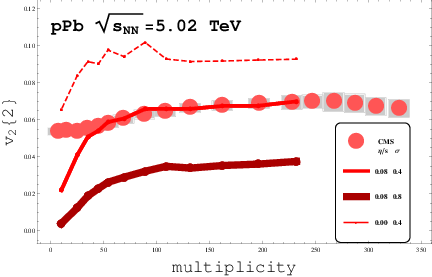}
 \caption{ (Color online) 
Integrated charged hadron $v_2\{2\}$ in 5.02 GeV p-Pb collisions for $|\eta|$ < 2.4 and $p_T$ > 0.3 GeV compared to measurements from the CMS Collaboration \cite{Chatrchyan:2013nka}.}  
\label{fig:intv2}
\end{figure}

\begin{figure}
 \includegraphics[width=\linewidth]{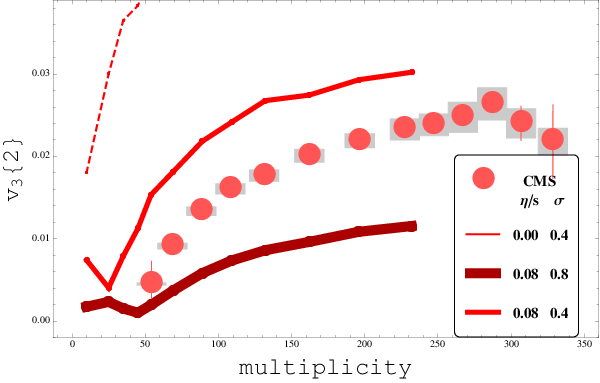}
 \caption{ (Color online) 
Integrated charged hadron $v_3\{2\}$ in 5.02 GeV p-Pb collisions for $|\eta|$ < 2.4 and $p_T$ > 0.3 GeV compared to measurements from the CMS Collaboration \cite{Chatrchyan:2013nka}.}  
\label{fig:intv3}
\end{figure}

The momentum of each particle can be varied independently, and an entire (symmetric) matrix can be formed.  Alternatively, one or both particles can be unrestricted, leading to an average over part of the matrix.  This is how the usual flow measurements are formed.  In particular, the momentum-integrated two-particle cumulant is obtained when neither particle is restricted to a particular bin, and the matrix is therefore an average over $p_T$
\begin{align}
\bar{v}_n\{2\} & \equiv \sqrt{V_{n\Delta}(\bar p_T, \bar p_T)} \\
& \stackrel{\rm flow}{=} \sqrt{\langle \bar v_n^2 \rangle} .
\label{vn2hydro}
\end{align}
The second line shows how the observable depends on the single-particle distribution (Eq.~\ref{vn}, integrated over $p_T$) in a hydrodynamic calculation \cite{Luzum:2013yya}.
The results of our hydrodynamic calculation for $\bar v_2\{2\}$ and $\bar v_3\{2\}$ are compared to experimental data in Figs.~\ref{fig:intv2} and \ref{fig:intv3}.  Viscosity has the expected effect of suppressing $v_n$.   Increasing $\sigma$ causes a decrease in the spatial eccentricity, and therefore also has a suppressing  effect.

\begin{figure}
 \includegraphics[width=\linewidth]{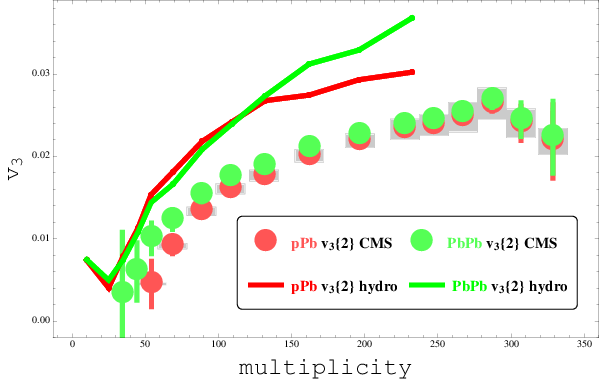}
 \caption{ (Color online) 
Integrated charged hadron $v_3\{2\}$ for the parameter set $\sigma$ = 0.4 rm and $\eta/s$ = 0.08 in p-Pb compared to Pb-Pb collisions \cite{Chatrchyan:2013nka}.}  
\label{fig:v3PbPb}
\end{figure}

An interesting feature that was noticed is that if one plots $v_3\{2\}$ from p-Pb and Pb-Pb collisions as a function of multiplicity (i.e., comparing high-multiplicity events in p-Pb to peripheral events in Pb-Pb with the same multiplicity), the results are very similar \cite{Chatrchyan:2013nka}.   It has been questioned whether this is natural (or even possible) in a hydrodynamic picture.  We find that even in our simple model, this result is approximately reproduced (see Fig.~\ref{fig:v3PbPb}).

\begin{figure}
 \includegraphics[width=\linewidth]{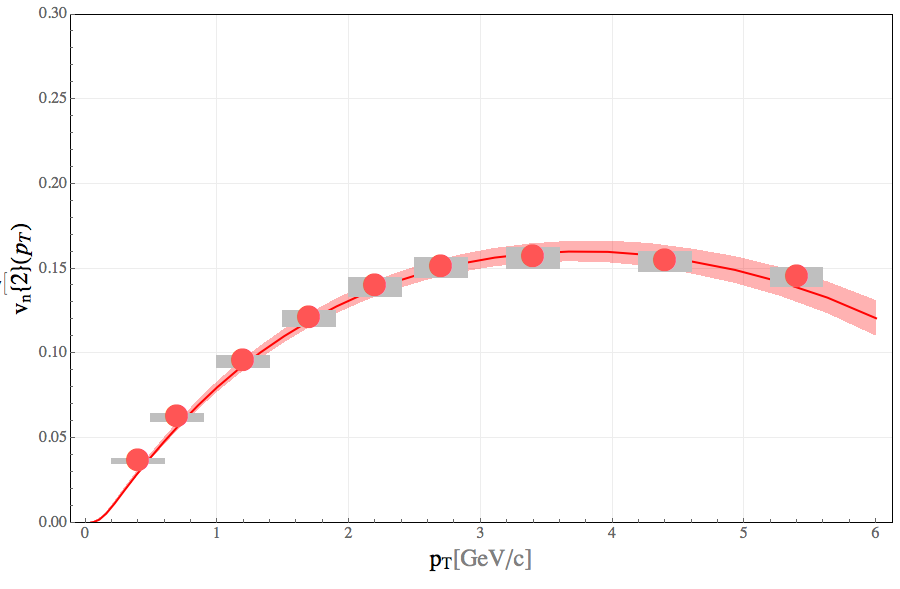}
 \caption{ (Color online) 
Differential charged hadron $v_2\{2\}$ in 5.02 GeV p-Pb collisions corresponding to the most central bin of Table \ref{tab} \cite{Chatrchyan:2013nka}.}  
\label{fig:diffv2}
\end{figure}

One can study the dependence on transverse momentum with a differential $v_n$ measurement
\begin{align}
v_n\{2\}(p_T) &\equiv \frac{V_{n\Delta}(p_T, \bar p_T)} {\bar v_n\{2\}} \\
& \stackrel{\rm flow}{=} \frac {\left\langle v_n(p_T)\ \bar v_n \cos n\left(\Psi_n(p_T) - \bar\Psi_n\right) \right\rangle} {\sqrt{\langle \bar v_n^2 \rangle}}
\end{align}
Here, $v_n(p_T)$ and $\Psi_n(p_T)$ are defined in each hydro event according to Eq.~\eqref{vn}, while $\bar v_n$ and $\bar \Psi_n$ are again defined by the analogous equation integrated over $p_T$.
We show $v_2\{2\}(p_T)$ for the highest multiplicity bin in Fig.~\ref{fig:diffv2}.

More information can be obtained by computing correlations between more than two particles.  Of particular interest is the four particle cumulant
\begin{align}
\label{vn4}
\bar v_n\{4\}^4 \equiv& -\left\langle \frac 1 {N_{\rm quad.}} \sum_{\rm quad.} \cos n(\phi_1 + \phi_2 - \phi_3 - \phi_4) \right\rangle \nonumber\\
& + 2 \bar v_n\{2\}^4 \\
\stackrel{\rm flow}{=} & \ 2 \langle \bar v_n^2 \rangle^2 - \langle \bar v_n^4 \rangle .
\label{vn4hydro}
\end{align}
The second line again indicates the quantity calculated in our hydrodynamic model.  If the quantity is positive, the observable $\bar v_n\{4\}$ is obtained by taking the fourth root.

\begin{figure}
 \includegraphics[width=\linewidth]{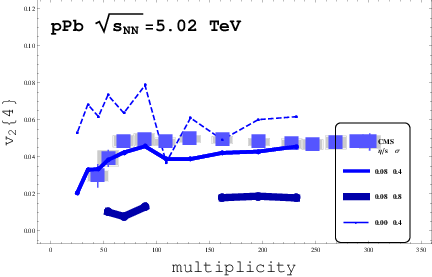}
 \caption{ (Color online) 
Integrated charged hadron $v_2\{4\}$ in 5.02 GeV p-Pb collisions \cite{Chatrchyan:2013nka}.}  
\label{fig:intv24}
\end{figure}

Because of how the two-particle correlation is subtracted from a four-particle correlation, non-flow correlations are typically suppressed. As a result, calculations that do not assume strong collective behavior typically give a value very close to zero, while hydrodynamic calculations typically predict a sizable value.  Remarkably, high-multiplicity p-Pb data show a large value that is clearly compatible with a hydrodynamic picture, as shown in Fig.~\ref{fig:intv24}.  To date, only hydrodynamic calculations have been able to reproduce this large value of $v_2\{4\}$.

\begin{figure}
 \includegraphics[width=\linewidth]{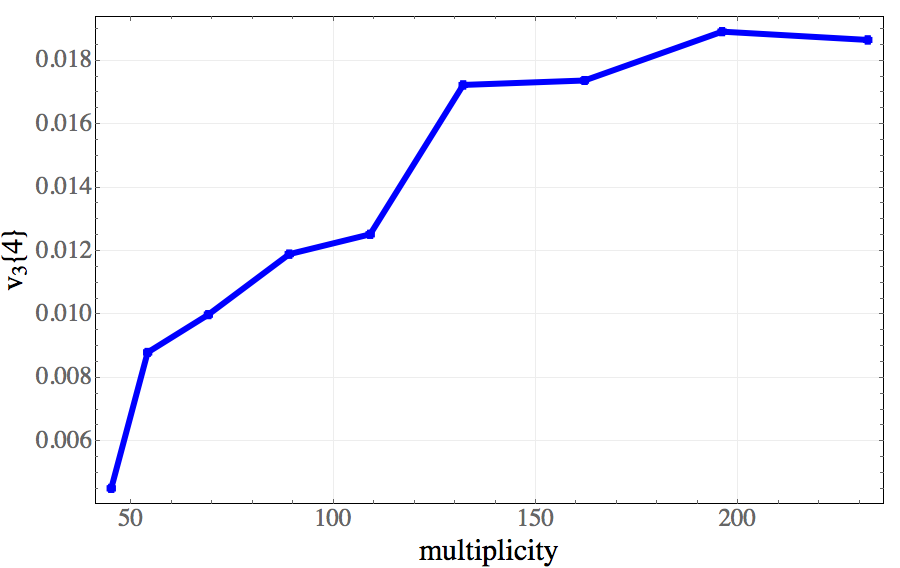}
 \caption{ (Color online) 
Predicted integrated charged hadron $v_3\{4\}$ in 5.02 GeV p-Pb collisions for $\sigma$ =  0.4 fm, $\eta/s$ = 0.08, and $p_T > $ 0.3 GeV}.
\label{fig:intv34}
\end{figure}

We also include a prediction for $\bar  v_3\{4\}$ in Fig.~\ref{fig:intv34}.  A measurement of this quantity could serve as a further non-trivial test of the nature of the collision system.


\section{A Stricter Test of Hydrodynamics}
%
%
It is clear from these results that a hydrodynamic model can in principle fit existing data.  Even within the simple model presented here, a slight tuning of parameters might well give a reasonable description of the available measurements.  Thus, the hydrodynamic picture is confirmed as plausible.  
One might wonder if there exists an even more stringent test that one could perform to test this picture.  For example, could a measurement be made that returns a value that a hydrodynamic calculation could never reproduce, no matter what parameters are chosen or what model of initial condition is used?  In fact, this is indeed possible.

The essential aspect of a purely hydrodynamic description is that particles emerge from the fluid independently --- that is, the single-particle distribution \eqref{vn} contains all relevant information, and multiparticle distributions are simply obtained as products of the single-particle distribution.  It turns out that this places non-trivial restrictions on observed multi-particle measurements, even if one is free to choose an arbitrary single-particle distribution.

As a simple example, by comparing Eqs.~\eqref{vn2hydro} and \eqref{vn4hydro}, it is clear that in any hydrodynamic calculation, one must have $v_n\{4\}^4 \leq v_n\{2\}^4$.  This does not serve as a non-trivial restriction, however, because it is generically true in most reasonable physical models.

In fact, no single piece of data presented above has a non-trivial restriction --- while a simultaneous description of all the data in a realistic model is non-trivial, in principle a specially engineered hydrodynamic solution could likely accommodate a large range of values.

However, one should note that not all possible information is being utilized --- part of the two-particle correlation matrix $V_{n\Delta}$ is being averaged over, even in a `differential' flow measurement $v_n\{2\}(p_T)$.  It turns out that the full matrix contains more useful information.

First recall that in any hydrodynamic calculation, the correlation matrix will have the form \cite{Gardim:2012im}
\begin{equation}
V_{n\Delta} (p_T^a, p_T^b)  \stackrel{\rm flow}{=} \left\langle v_n(p_T^a) v_n(p_T^b) \cos n \left[\Psi_n(p_T^a) - \Psi_n(p_T^b) \right] \right\rangle
\end{equation}
That is, it's the event average of a scalar product between the flow vector at each of the two values of $p_T$.  This form dictates that the elements of the matrix must satisfy a set of inequalities \cite{Gardim:2012im}.  

First, the diagonal elements (that is, when both particles are restricted to the same $p_T$ bin $a$) must be positive semidefinite
\begin{equation}
V_{n\Delta} (p_T^a, p_T^a) \geq 0
\end{equation}
The off-diagonal elements must be related to the diagonal by a triangle inequality
\begin{equation}
V_{n\Delta} (p_T^a, p_T^b)^2 \leq V_{n\Delta} (p_T^a, p_T^a) V_{n\Delta} (p_T^b, p_T^b) .
\end{equation}

We reiterate that these inequalities are inescapable.  They will be satisfied in any purely hydrodynamic calculation, and can not be circumvented by engineering a particular initial condition or tuning parameters.  It should also be noted that they are not trivial.  Recall that in Pb-Pb collisions, these inequalities are satisfied everywhere hydrodynamics is expected to be valid, but broken everywhere else \cite{Ollitrault:2012cm} -- that is, the first inequality is broken at high $p_T$ (specifically in the third harmonic $V_{3\Delta}$), while the second is broken at all momenta in the first harmonic (since it has long been known that $V_{1\Delta}$ should have a contribution from a correlation due to momentum conservation, and it indeed is consistent with  a combination of flow and momentum conservation \cite{Retinskaya:2012ky}).

We first propose that the entire double differential correlation matrix be measured and compared against these inequalities as a stringent test of the hydrodynamic picture.  Any violation unambiguously indicates a breakdown of hydrodynamics as the dominant contribution to correlations, and the presence of at least some significant contribution from non-flow correlations.

Whenever the first inequality is satisfied, a convenient way to quantitatively compare results is to create the ratio \cite{Gardim:2012im}
\begin{equation}
\label{correlation}
r_n\equiv\frac{V_{n\Delta}(p_T^a,p_T^b)}{\sqrt{V_{n\Delta}(p_T^a,p_T^a)V_{n\Delta}(p_T^b,p_T^b)}} .
\end{equation}
In hydrodynamics, the second inequality ensures that this ratio must lie in the range $-1 \leq r_n \leq 1$.  
A value of 1 indicates that the inequality is saturated.  This is the case only in the limit where there are no event-by-event flow fluctuations.  Stronger fluctuations will tend to drive it further from one.

However, if hydrodynamics is not the correct description, the quantity is completely unbounded.  While $V_{n\Delta}$ is bounded by $\pm 1$, any value is allowed for $r_n$, and even existing data for differential flow $v_n\{2\}(p_T)$ does not restrict the possible values for $r_n$.  So this measurement would provide a significant additional constraint to theoretical models, beyond what has already been measured.

\begin{figure}
 \includegraphics[width=\linewidth]{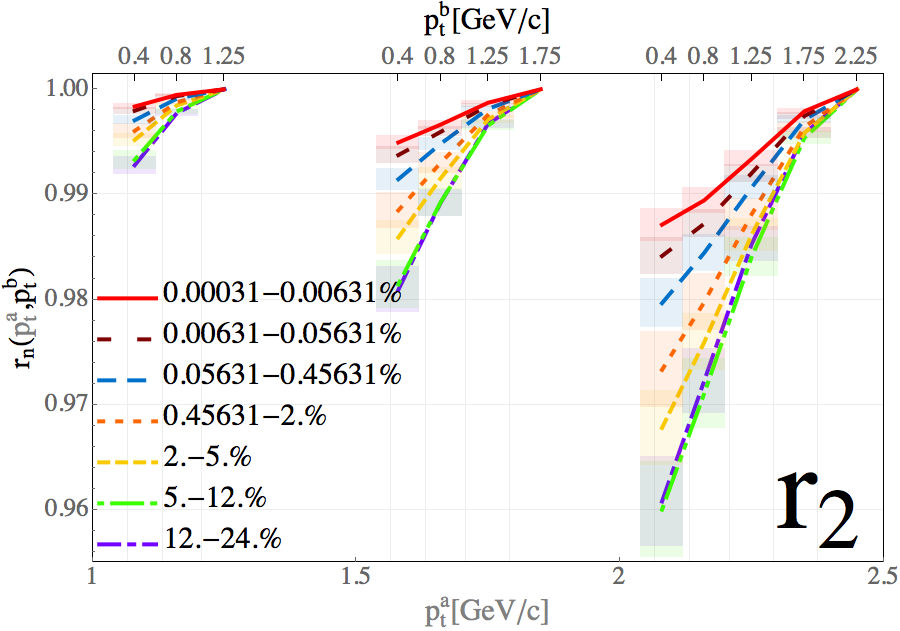}
 \caption{ (Color online) 
$r_2$ for the 7 highest multiplicity bins in Table \ref{tab}.  The 2D matrix is flattened along the x-axis, such that each set of curves represents a fixed transverse momentum bin for one of the particles ($p_T^a$, labeled on the bottom), while each point on a curve represents a different $p_T^b$  We predict that the highest multiplicity collisions should give a value close to 1, and decreasing monotonically with decreasing multiplicity.}  
\label{fig:r2cent}
\end{figure}

\begin{figure}
 \includegraphics[width=\linewidth]{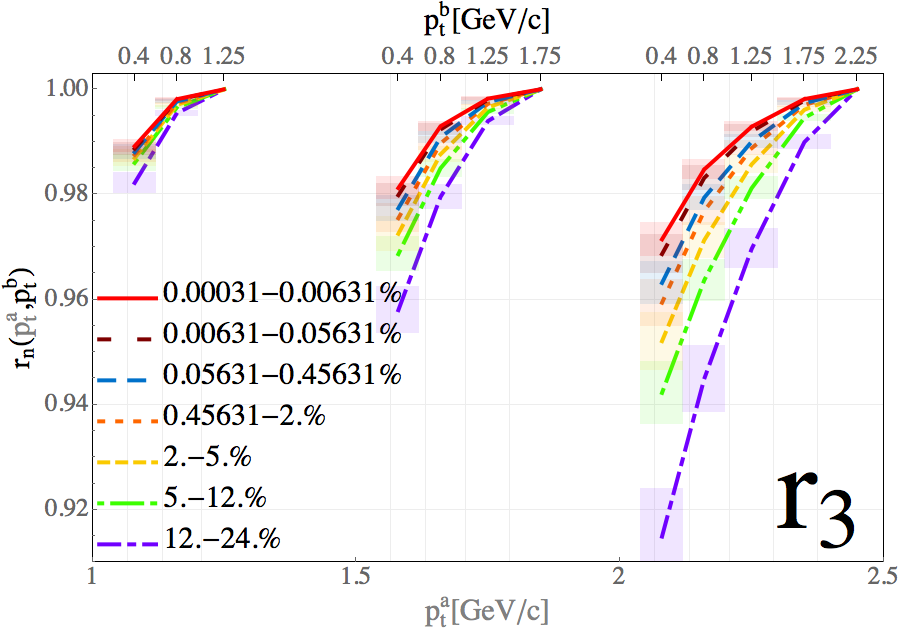}
 \caption{ (Color online) 
As for Fig.~\ref{fig:r2cent}, for the third azimuthal harmonic $r_3$.  We predict the same qualitative trend with multiplicity.}
\label{fig:r3cent}
\end{figure}

While any value between -1 and 1 are allowed mathematically in hydrodynamics, our calculations indicate that in the highest-multiplicity collisions, the value should be expected to be very close to 1, which gets farther from one as multiplicity decreases.  In Figs.~\ref{fig:r2cent} and \ref{fig:r3cent} we show the centrality dependence of $r_2$ and $r_3$, respectively, for the set of parameters that best fit the above data ($\sigma = 0.4 fm$, $\eta/s = 0.08$), though the trend is general.  
Thus, any deviation from this trend would likely indicate a breakdown of a dominantly fluid description.


\begin{figure}
 \includegraphics[width=\linewidth]{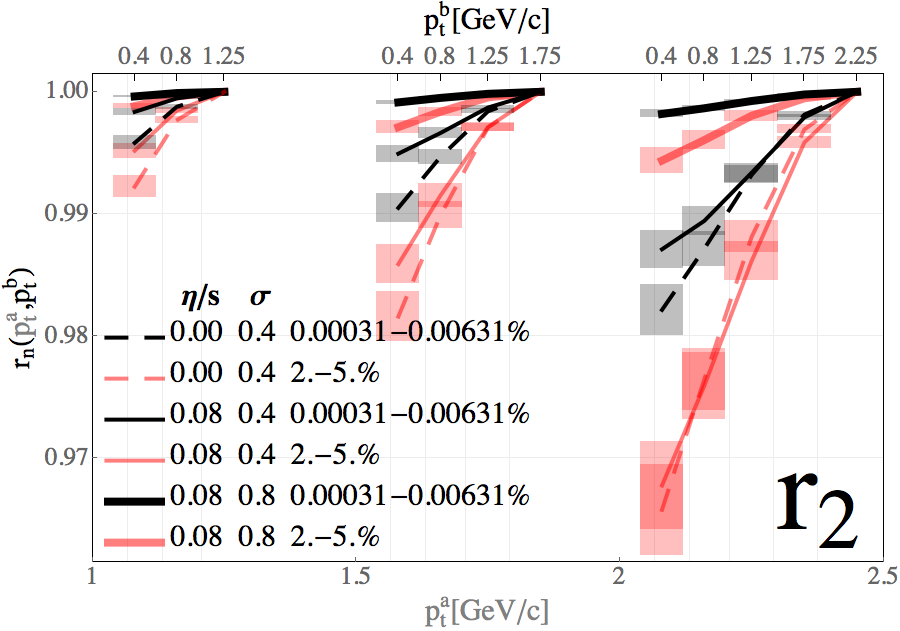}
 \caption{ (Color online) 
$r_2$ for 2 centrality bins, with varying values of $\eta/s$ and $\sigma$ (in fm).}  
\label{fig:r2sigmaeta}
\end{figure}

\begin{figure}
 \includegraphics[width=\linewidth]{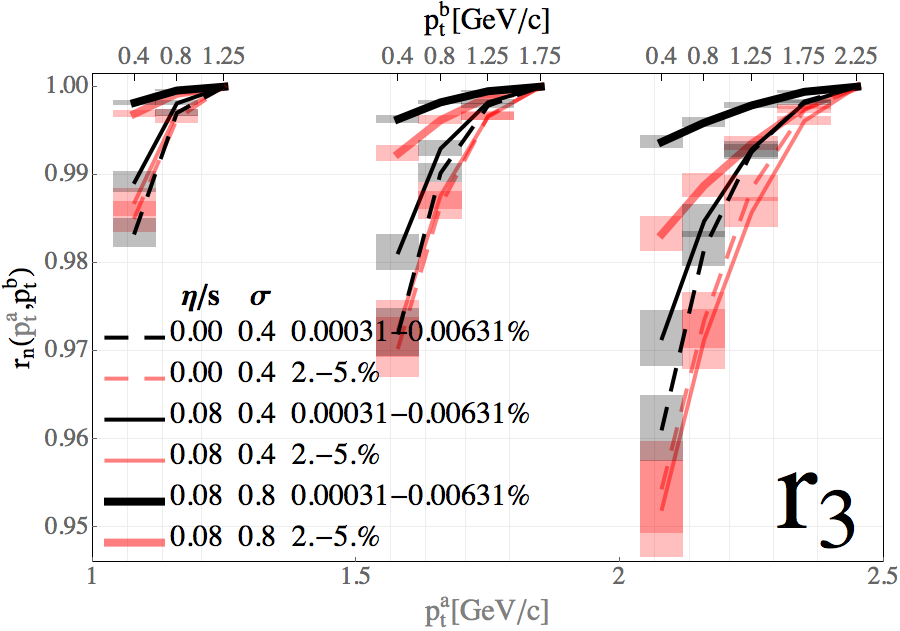}
 \caption{ (Color online) 
$r_3$ for 2 centrality bins, with varying values of $\eta/s$ and $\sigma$ (in fm).}  
\label{fig:r3sigmaeta}
\end{figure}

In Figs.~\ref{fig:r2sigmaeta} and \ref{fig:r3sigmaeta}, we show the effect of varying the viscosity $\eta/s$ and the  granularity parameter $\sigma$.  We find that the viscosity actually has a quite small effect.  In contrast, the stronger affect of varying $\sigma$ indicates that aspects of the initial condition much more important.  
We confirmed that this is still the case in mid-peripheral Pb-Pb collisions.  This could therefore be a potentially useful observable, since numerous other observables are sensitive to $\eta/s$, but no other observable is known to be this sensitive to the granularity of the initial conditions of a heavy-ion collision.
\section{Revisiting $r_n$ for A-A}
Three previous hydrodynamic calculations exist of the ratio $r_n$, all for heavy ion collisions.  The first was a calculation for Au-Au collisions with NeXus initial conditions and zero viscosity \cite{Gardim:2012im}.  The next, in a Pb-Pb system, was a Glauber model calculation with $\eta/s = 0.08$ (which we will refer to as the second calculation) and an MC-KLN calculation with $\eta/s = 0.2$ \cite{Heinz:2013bua}.

The values of $r_n$ in these three calculations were ordered according to the viscosity in the calculation --- the lowest viscosity result was farthest from one, while the calculation with the largest viscosity was closest to one.  Both of the latter two had reasonable agreement with measured Pb-Pb data.  The most natural interpretation was that viscosity reduced the magnitude of fluctuations and causes the ratio to become closer to 1.  However, one should note that each of these three calculations also had different initial conditions.  As we have shown, viscosity actually has a very small affect, while the initial conditions can have a significant effect.  The latter two calculations neglect the local fluctuations in entropy production.  While these are typically not believed to be important for observables in heavy-ion collisions, we might guess that it is these extra fluctuations that cause $r_n$ to deviate farther from one. 

\begin{figure}
 \includegraphics[width=\linewidth]{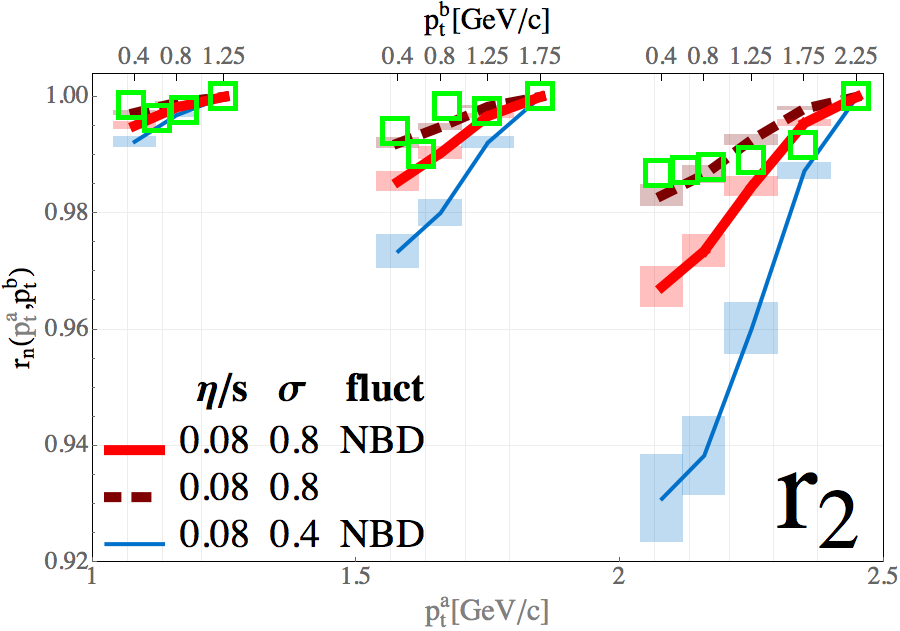}
 \caption{ (Color online) 
$r_2$ in 40--50\% centrality Pb-Pb collisions from our model with and without negative binomial fluctuations (NBD) in the initial entropy per participant and with different values of the granularity parameter $\sigma$ (in fm), compared to data derived from ALICE measurements of $V_{2\Delta}$ \cite{Aamodt:2011by, Gardim:2012im}.}  
\label{fig:r2PbPb}
\end{figure}

\begin{figure}
 \includegraphics[width=\linewidth]{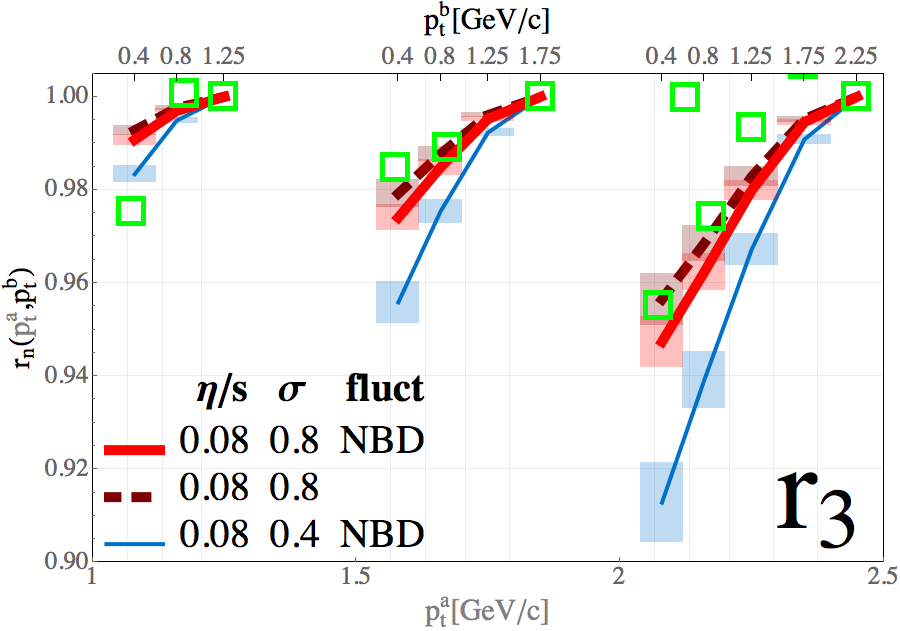}
 \caption{ (Color online) 
$r_3$ in 40--50\% centrality Pb-Pb collisions from our model with and without negative binomial fluctuations (NBD) in the initial entropy per participant and with different values of the granularity parameter $\sigma$ (in fm), compared to data derived from ALICE measurements of $V_{2\Delta}$ \cite{Aamodt:2011by, Gardim:2012im}.}  
\label{fig:r3PbPb}
\end{figure}

To test this, we calculated $r_n$ at 40-50\% centrality Pb-Pb in the same model as above, but in addition we did a calculation for the same events with the extra NBD fluctuations turned off (i.e., so that every participant had the same contribution to the total entropy).  The results are shown in Figs.~\ref{fig:r2PbPb} and \ref{fig:r3PbPb}.

Our calculations with a more standard Glauber model and $\sigma = 0.8$ fm should be close to the previous results, which fit data well.  As can be seen in Figs.~\ref{fig:r2PbPb} and \ref{fig:r3PbPb}, this is the case.  However, when we add more realistic negative binomial fluctuations, the value decreases, away from data.  
Further, if we change to the model that best fits existing p-Pb data ($\sigma = 0.4$ fm), $r_2$ falls even farther from the measured value.  This indicates that a simultaneous fit to all data may be more difficult than currently believed, and the tension with data may ultimately provide very strict constraints on hydrodynamic models. 

\section{Summary}
We have shown that a hydrodynamical model can plausibly describe a wide range of observables that have been measured in high multiplicity proton-lead collisions.  In order to make a more stringent test of whether this is the correct description, we propose to measure a new observable $r_n$, which uses the full transverse momentum information from two-particle correlations.  This has the potential to rule out hydrodynamics as the correct description.  Further, while it does not serve as a useful probe of viscosity, it is a promising observable to probe the transverse length scale of fluctuations in the early stages of a heavy-ion collision.
\begin{acknowledgments}
%
This work was funded in part by the Natural Sciences and Engineering Research Council of Canada, and in part by the Office of Nuclear Physics in the US Department of EnergyÕs Office of Science under Contract No. DE-AC02-05CH11231. I. K. acknowledges support through a fellowship from the Canadian Institute of Nuclear Physics, G. S. D. acknowledges support through a Banting Fellowship from the Natural Sciences and Engineering Research Council of Canada, and C.G. acknowledges support from the Hessian Initiative for Excellence (LOEWE) through the Helmholtz International Center for FAIR (HIC for FAIR).
\end{acknowledgments}

\end{document}